# Tunable polymorphism of epitaxial iron oxides in the four-in-one ferroic-on-GaN system with magnetically ordered α-, γ-, ε-$Fe_2O_3$ and $Fe_3O_4$ layers

*S. M. Suturin*[1], A. M. Korovin[1], S.V. Gastev[1], M. P. Volkov[1], M. Tabuchi[2], N. S. Sokolov[1]

[1] Ioffe Institute, 26 Polytechnicheskaya str., St. Petersburg, 194021 Russia

[2] Synchrotron Radiation Research Center, Nagoya University, Furo-cho, Chikusa, Nagoya 464-8603, Japan

* Correspondence e-mail: suturin@mail.ioffe.ru

Hybridization of semiconducting and magnetic materials into a single heterostructure is believed to be potentially applicable to the design of novel functional spintronic devices. In the present work we report epitaxial stabilization of four magnetically ordered iron oxide phases ($Fe_3O_4$, γ-$Fe_2O_3$, α-$Fe_2O_3$ and most exotic metastable ε-$Fe_2O_3$) in the form of nanometer sized single crystalline films on GaN(0001) surface. The epitaxial growth of as many as four distinctly different iron oxide phases is demonstrated within the same single-target Laser MBE technological process on a GaN semiconductor substrate widely used for electronic device fabrication. The discussed iron oxides belong to a family of simple formula magnetic materials exhibiting a rich variety of outstanding physical properties including peculiar Verwey and Morin phase transitions in $Fe_3O_4$ and α-$Fe_2O_3$ and multiferroic behavior in metastable magnetically hard ε-$Fe_2O_3$ ferrite. The physical reasons standing behind the nucleation of a particular phase in an epitaxial growth process deserve interest from the fundamental point of view. The practical side of the presented study is to exploit the tunable polymorphism of iron oxides for creation of ferroic-on-semiconductor heterostructures usable in novel spintronic devices. By application of a wide range of experimental techniques the surface morphology, crystalline structure, electronic and magnetic properties of the single phase iron oxide epitaxial films on GaN have been studied. A comprehensive comparison has been made to the properties of the same ferrite materials in the bulk and nanostructured form reported by other research groups.

## I. INTRODUCTION

Hybrid heterostructures combining closely spaced semiconducting and magnetic layers are believed to be promising candidates to be used in functional spintronic devices. Unlike diluted magnetic semiconductors, the hybrid heterostructures allow separate control over the magnetic and electrical properties. The magnetic layers in such heterostructures must preferably be dielectric or have controllable conductivity as the high concentration of free charge carriers at the interface would provide an additional non-radiative recombination channel leading to significant deterioration of the semiconducting properties. Transparency of the magnetic layer in the visible range would be another demand as long as the optoelectronic features of the semiconducting device are to be exploited. Making choice of the magnetic layer material for the hybrid heterostructure one could consider iron oxides as suitable candidates. Including FeO, $Fe_3O_4$ and a number of $Fe_2O_3$ polymorphs the iron oxides make up a big family of magnetic materials exhibiting a rich variety of outstanding physical properties, presenting interest both for technological applications and fundamental studies. Below a short description of the four iron oxides discussed in this paper will be given with stress put on the very different structural, magnetic and electric properties of these materials.

Magnetite ($Fe_3O_4$) is crystallized in the cubic inverse spinel lattice (Fd-3m space group, a=8.398 Å). The easy magnetization axis lies along the [111] direction. Having the largest magnetic moment among the iron oxides (2.66 $\mu_B$ / formula unit (f.u.), $M_s$=480 emu/cm$^3$, $4\pi M_s$ = 6000 G) and Néel temperature of 850 K $Fe_3O_4$ has been long used in magnetic recording applications. Magnetite is a half-metal with full spin polarization at the Fermi level [1]. Below the Verwey transition temperature of 120 K the conductivity and magnetization in high purity magnetite drops by several orders of magnitude due to the charge ordering in the Fe tetrahedral sites. The ferromagnetic order in magnetite comes from the iron magnetic moments: parallel to the magnetic field in octahedral sites and antiparallel in tetrahedral sites.

Maghemite ($\gamma$-$Fe_2O_3$) can be considered as fully oxidized magnetite and crystallizes in the same inverse spinel lattice (a=8.330 Å). The $Fe^{III}$ vacancy distribution in octahedral positions gives rise to several possible crystal symmetries: the random distribution (space group Fd-3m as in $Fe_3O_4$), partially ordered distribution (space group $P4_132$ as in $LiFe_5O_8$) and ordered distribution (space group $P4_32_12$). Maghemite exhibits ferrimagnetic order with a high Néel temperature of 950 K. Being insulating and having a magnetic moment of 2.5 $\mu_B$ / f.u. ($M_s$=380 emu/$cm^3$, $4\pi M_s$ = 4800 G) it is supposed a good candidate to act as a tunnel barrier for the spin filtering devices [2]. Maghemite is also widely used in magnetic storage, gas sensing and biomedical applications [3–5].

Hematite ($\alpha$-$Fe_2O_3$) is the most thermodynamically stable iron oxide crystallizing in the R-3c:H space group (a=5.0355 Å, c=13.7471 Å). Between Morin transition temperature of 250 K and Neel temperature of 950 K the low symmetry $Fe^{III}$ sites allow spin–orbit coupling to cause canting of the moments lying in basal plane perpendicular to the c axis. The corresponding magnetic moment is very small (<0.02 $\mu_B$ / f.u., $M_s$<2 emu/$cm^3$, $4\pi M_s$ < 30 G). Below the Morin transition the anisotropy makes the moments align antiferromagnetically along the c axis. This transition is sometimes suppressed in the nanoscale objects as well as in the bulk crystals due to the presence of impurities and defects.

Metastable epsilon ferrite ($\varepsilon$-$Fe_2O_3$) iron oxide receiving much attention nowadays due to its exotic properties has an orthorhombic noncentrosymmetric lattice ($Pna2_1$ space group, a=5.089 Å, b=8.780 Å, c=9.471 Å) [6] isostructural to $GaFeO_3$ [7], $AlFeO_3$ [8], $kGa_2O_3$ [9] and $kAl_2O_3$ [10]. Iron atoms in $\varepsilon Fe_2O_3$ occupy four nonequivalent crystallographic sites, including one tetrahedral ($T_d$) site and three octahedral ($O_h$) sites. The uncompensated magnetic moments of distorted $T_d$ and undistorted $O_h$ sublattices account for ferrimagnetic behavior of $\varepsilon Fe_2O_3$ below $T_N$=490-500 K with resultant magnetization of 0.6 $\mu_B$ / f.u. at RT ($M_s$=100 emu/$cm^3$, $4\pi M_s$ = 1300 G) [11]. The distortions of the two $O_h$ and one $T_d$ sites explain the large orbital moment leading to significant spin-orbit coupling [11], high magnetocrystalline anisotropy (K=4·$10^6$ erg / $cm^3$) and very high coercivity (above 2 T) [12]. The epsilon ferrite exhibits room temperature multiferroic properties [13], magnetoelectric coupling [14] and millimeter wave 190 GHz absorption at room temperature [15,16]. Because of $\varepsilon Fe_2O_3$ metastability most works deal with randomly oriented nanoparticles [17–19]. Only few works report epitaxial growth of $\varepsilon$-$Fe_2O_3$ layers on STO, $Al_2O_3$ and YSZ using $GaFeO_3$ or $AlFeO_3$ transition layers [13,20–22].

Due to numerous applications and intrinsic magnetic properties the fabrication of various iron oxides in nanoscale form has been widely studied during the past decades. Growth of iron ferrite nanoparticles has been performed by a variety of methods including laser pyrolysis, co-precipitation, sol–gel, microemulsion and ball-milling [3]. Epitaxial layers have been grown on different substrates (InAs, GaAs, $MgAl_2O_4$, Si, etc.) using reactive oxygen or ozone assisted molecular beam epitaxy, reactive magnetron sputtering and pulsed laser deposition. Surprisingly there are very few works describing iron oxide hybridization with GaN semiconductor widely utilized in modern high-power electronics. Growth of various other oxides ($SiO_2$, $Al_2O_3$, $HfO_2$, $Ga_2O_3$) on GaN has been the object of study in search of appropriate gate insulators for metal-oxide-semiconductor devices [23–26]. The advantage of iron oxides is that they feature controllable spontaneous magnetization/polarization that may add new functionality to the GaN-based semiconductor devices, e.g. can be used for impedance matching spin injection into GaN (from half metallic $Fe_3O_4$ or via tunneling through insulating $\gamma$-$Fe_2O_3$) or for ferroelectric gate insulators. The GaN-based spin relaxation measurements have shown spin lifetimes of 20 ns at 5 K [27] while theoretical calculations predict that spin lifetime in pure GaN is about three orders of magnitude larger than in GaAs [28]. The examples of using iron oxides for spin filtering (though not in combination with GaN) are discussed in [2] where epitaxial γ-$Fe_2O_3$ films were fabricated on Nb:$SrTiO_3$. Of various iron oxides only $Fe_3O_4$ has been ever grown on GaN [29,30] by combining room temperature deposition of iron in UHV with short post growth annealing in oxygen. The growth of $Fe_3O_4$ onto GaN by MOCVD using $Ga_2O_3$ buffer layer has been also reported [31].

Following our earlier brief communication[32], the present work reports the recent progress in epitaxial stabilization of the four structurally and magnetically different iron oxide phases ($Fe_3O_4$, α-$Fe_2O_3$, γ-$Fe_2O_3$ and the most exotic ε-$Fe_2O_3$) in the form of nanometer sized single crystalline films on GaN(0001) surface. After describing the experimental details in section II, the growth and identification of the four iron oxide phases will be discussed in detail in section III. Upon introducing the epitaxial technology and speaking about the surface morphology of the grown films, the common building principles of the iron oxide direct and reciprocal lattices will be addressed. Having settled a convenient coordinate system to work in, we will discuss the ways to identify the crystal structure and epitaxial relations in the studied system by using high energy electron diffraction 3D reciprocal space mapping. Section IV is dedicated to the complementary X-ray diffraction studies. Specular crystal truncation rods are analyzed to give accurate evaluation of the out-of-plane interlayer spacing in the stabilized iron oxide phases and the interface transition layer. X-ray absorption and X-ray magnetic circular dichroism are applied in section V to investigate oxidation state and coordination of iron atoms in the ferrimagnetically ordered sublattices of the studied oxides. A comparison to the properties of the corresponding bulk materials is carried out. Finally the in-plane magnetization reversal in $Fe_xO_y$ / GaN layers is described in section VI and the summary is given in section VII.

## II. EXPERIMENTAL

The few micrometer thick Ga terminated GaN layers acting as the host surface for the iron oxide deposition were grown in Ioffe Institute by means of MOVPE [33] on the $Al_2O_3$(0001) substrates. The iron oxide films having thickness of 10-60 nm were grown onto the GaN (0001) surface by means of laser molecular beam epitaxy. The α-$Fe_2O_3$ target was ablated by pulses of KrF excimer laser ($\lambda$ = 248 nm) at a fluence of 2-6 J/$cm^2$, the exact value having no immediate effect on the physical properties of the grown films. The iron oxide growth was performed in oxygen, nitrogen and argon atmosphere at pressures ranging from 0.02 to 0.5 mbar. The substrate temperature during iron oxide deposition was varied from room temperature (RT) to 850°C as measured inside the stainless steel sample holder to which the sample was clamped. As the GaN / $Al_2O_3$ substrates are transparent to the infra-red irradiation and the only heat transfer is through the direct contact to the sample holder, the real substrate temperature may be lower than measured.

For the real time control over the oxide layer lattice structure and orientation, an advanced 3D reciprocal space mapping by high-energy electron diffraction (RHEED) was used. The general idea of this approach was described in detail in our recent publication [34] where the incidence angle was varied during image acquisition. In the present work the much more informative variable-azimuth mapping has been applied with RHEED patterns collected during sample rotation around the surface normal [35]. Following this approach, the raw patterns (that are essentially spherical reciprocal space cuts) are used to build 3D intensity maps. The planar cuts and orthogonal projections of the resulting maps will be presented in this paper to describe the reciprocal space structure of the studied samples. The "side-views" (cuts performed perpendicular to the sample surface) are flattened improvements over the traditionally published raw patterns; the less conventional "plan-views" (projections along the lines parallel to the surface normal) cannot be visualized taking a single raw RHEED pattern and are very informative in understanding the in-plane reciprocal space structure especially when used to visualize the diffuse scattering features. The dedicated OpenGL software utilizing the extremely fast image processing capabilities of hundreds core modern graphic processing units has been developed by the authors to process the raw diffraction patterns. The developed software is capable of on-the-fly calculation of reciprocal space coordinates for clouds consisting of half billion pixels and building 3D intensity maps suitable for visualization and comparison to the model reciprocal lattices. To accommodate the wide range of RHEED map intensities within the 8 bit gray scale images, a background subtraction routine has been applied to the maps presented in this paper. It is worth noting that the main aim of the applied 3D mapping technique is to measure and model reflection positions and shapes rather than their absolute intensities as the latter are known to be subject to rather strong dynamical effects. The modelled reciprocal lattices shown on top of the maps take into account the systematic absences of the corresponding space groups. Though the extinction rules can be sometimes violated in electron diffraction due to multiple scattering effects (e.g. for bright many-spot patterns taken in symmetric azimuths), these effects can be easily ruled out by detuning from high symmetry azimuths as it is done in precession electron direction used in transmission electron microscopy.

The X-ray diffraction (XRD) measurements were carried out ex-situ at BL3A beamline of Photon Factory synchrotron (Tsukuba, Japan). The surface morphology was studied using ambient air NT-MDT atomic force microscope. The magnetic properties were measured at room temperature (RT) using a longitudinal MOKE setup described in [35] as well as using the Quantum Design PPMS VSM magnetometer at RT and at 100 K. The X-ray absorption (XAS) and X-ray circular magnetic dichroism (XMCD) studies have been carried out at RT at BL16 beamline of Photon Factory synchrotron (Tsukuba, Japan). To be sensitive to the in-plane magnetization the light incident angle was set to 30 deg while the magnetic field (up to 5 T) was applied parallel to the light propagation direction.

## III. GROWTH OF IRON OXIDES ON GALLIUM NITRIDE CONTROLLED BY RHEED RECIPROCAL SPACE 3D MAPPING

### A. Outline of the growth technology

In the present work the growth of four distinctly different $\alpha$-$Fe_2O_3$, $\gamma$-$Fe_2O_3$, $\varepsilon$-$Fe_2O_3$ and $Fe_3O_4$ iron oxides on GaN(0001) surface has been achieved through controlled variation of substrate temperature, deposition rate, background gas composition and pressure. The outline of the developed technology is sketched in Fig. 1. Three essentially distinct technological stages were used to grow iron oxides: growth in oxygen, annealing in oxygen and growth in nitrogen. Growth in oxygen results in formation of either $\alpha$-$Fe_2O_3$ phase at lower substrate temperature and pressure (T=400-600°C; p<0.02 mbar) or $\varepsilon$-$Fe_2O_3$ phase at elevated temperature and pressure (T=800°C; p=0.2-0.4 mbar). Growth in nitrogen (T=600-800°C @ 0.02-0.2 mbar) results in $Fe_3O_4$ phase stabilization. Finally the $\gamma$-$Fe_2O_3$ can be produced by exposing $Fe_3O_4$ layers to oxygen at 600°C-800°C.

As confirmed by the AFM measurements the iron oxides form uniform layers on the GaN step-and-terrace surface (Fig. 1 bottom). On the nanometer scale the film surface consists of mounds 15-20 nm in width and 1-3 nm in height. The flattest surface is achieved for the films grown at 600°C. Even in 40-60 nm films the step-and-terrace pattern inherited from the GaN surface can still be recognized indicating that surface roughness does not drastically increase with film thickness. The films grown at 800°C show a more pronounce height variation. The reasons for which the mounds are formed at the surface will be discussed later in this paper.

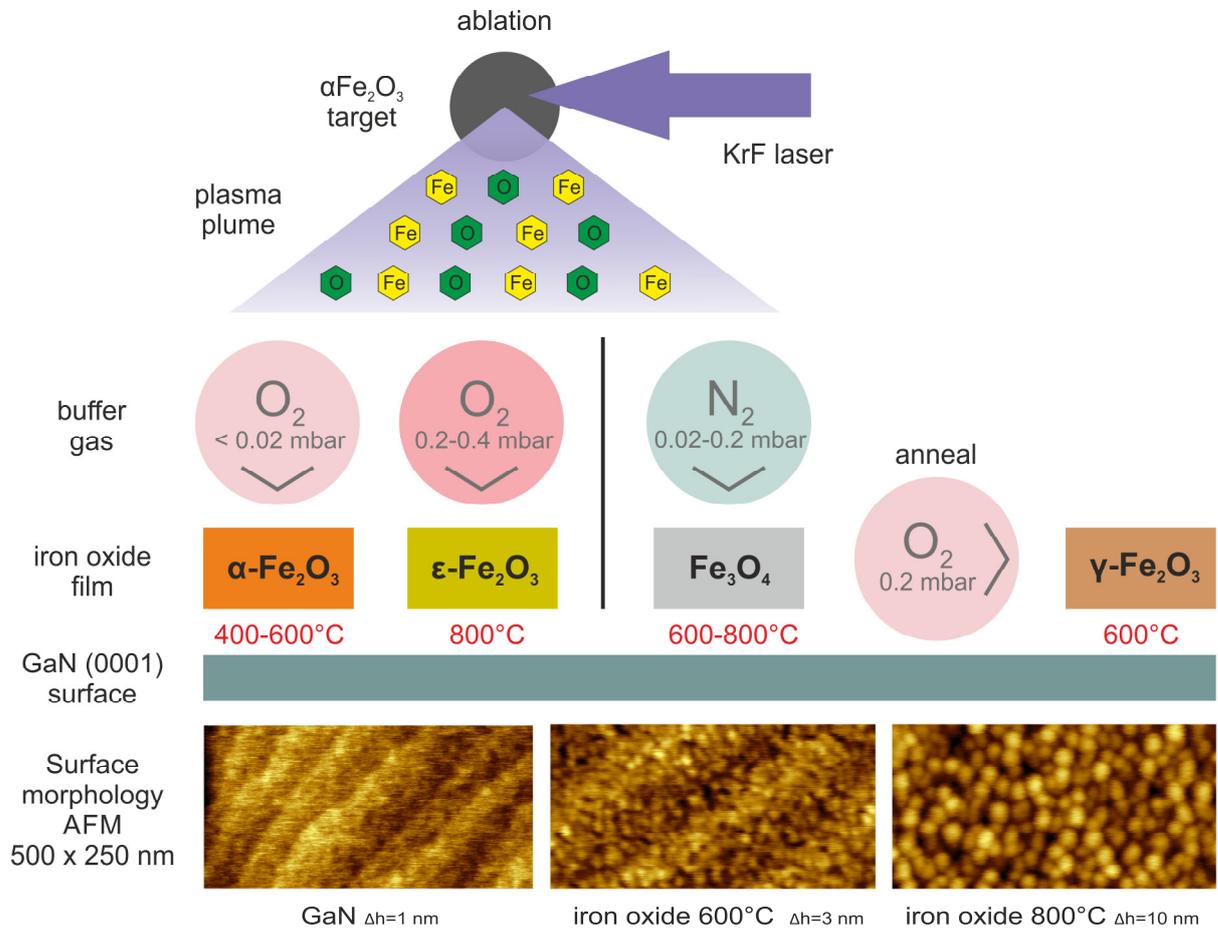

FIG. 1. The outline of the Laser MBE technology developed to grow epitaxial layers of various iron oxides on the GaN (0001) surface. The growth is performed by ablating a hematite target in $O_2$ or $N_2$ background atmosphere while the substrate is kept at 400-800°C. Post-growth annealing in oxygen is carried out to convert $Fe_3O_4$ films to $\gamma$-$Fe_2O_3$. AFM images show the step-and-terrace GaN surface as well as the typical columnar surface morphology of the iron oxide films grown at 600-800°C.

### B. Lattice structure of iron oxides and GaN - direct space

At first glance the trigonal $\alpha$-$Fe_2O_3$ (a=5.036 Å, c=13.747 Å), cubic $Fe_3O_4$ / $\gamma$-$Fe_3O_4$ / FeO (a=8.399 Å / 8.33 Å / 4.332 Å) and orthorhombic $\varepsilon$-$Fe_2O_3$ (a=5.089 Å, b=8.780 Å, c=9.471 Å) have very different lattice symmetries and drastically non-coincident lattice parameters. However at closer look the building principles of the four lattices appear to be rather similar - as shown in Fig. 2 (a) they consist of alternating oxygen and iron layers. The oxygen planes are arranged in slightly distorted (see Fig. 2b) close packed sequences with AB-AB stacking in $\alpha$-$Fe_2O_3$, ABC-ABC stacking in $\gamma$-$Fe_2O_3$ / $Fe_3O_4$ / FeO and ABCB-ABCB stacking in $\varepsilon Fe_2O_3$. The oxygen planes though differently indexed exhibit very similar interplane distances: 2.424 Å, 2.405 Å and 2.50 Å between the (111) planes in $Fe_3O_4$, $\gamma$-$Fe_2O_3$ and FeO; 2.368 Å between the $\varepsilon$-$Fe_2O_3$ (001) planes and 2.291 Å between the $\alpha$-$Fe_2O_3$ (0001) planes. The four oxides also differ in the way the iron atoms populate the slightly distorted tetrahedral and octahedral sites between the oxygen planes (see Fig. 2(a)). As will be shown below the iron oxide growth direction is perpendicular to the oxygen planes, thus the phase choice is controlled simply by the way how the oxygen and iron atoms are stacked upon deposition.

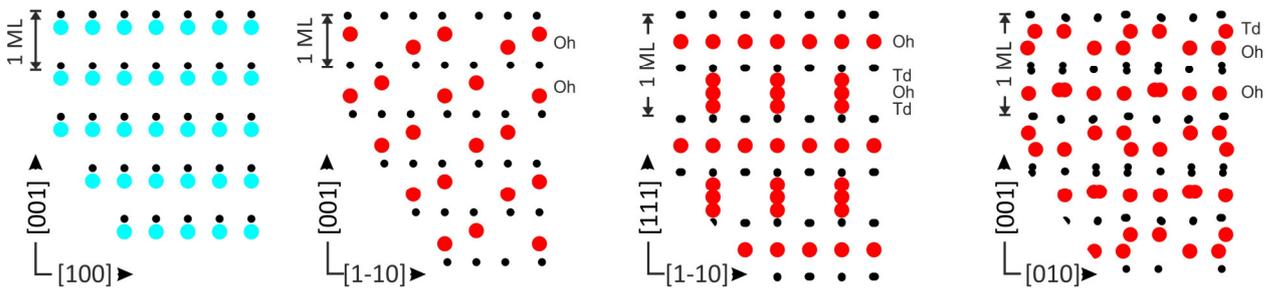

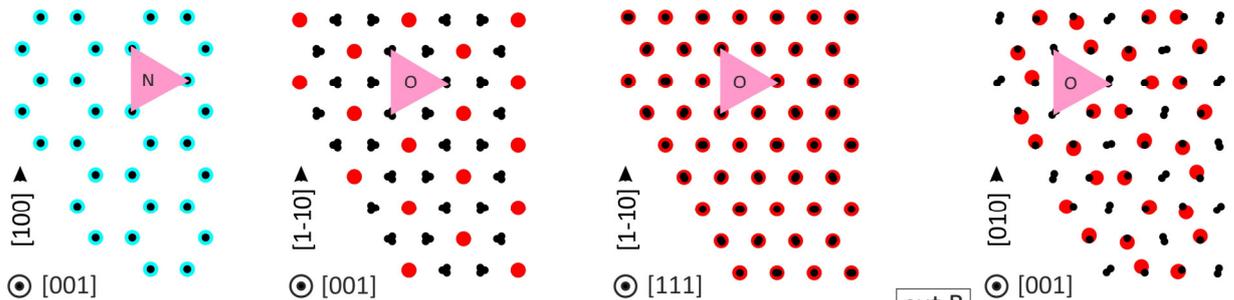

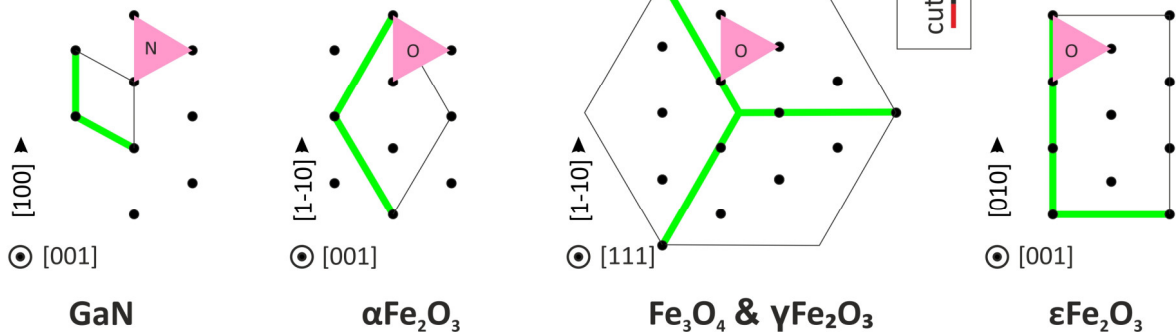

FIG. 2. Epitaxial relations in the iron-oxide-on-GaN system. Side view lattice projections (a) show the stacking sequence of Ga-N planes in gallium nitride and Fe-O planes in iron oxides. Plan view lattice projections (b) highlight the slightly distorted hexagonal arrangement of Fe-O atoms in the iron oxides. Plan view lattice cuts (c) show single anion planes together with the corresponding bulk unit cell vectors. Here and below projections represent the bulk lattice viewed along a certain direction, while cuts are lattice cross-sections containing just one plane of atoms or reciprocal lattice nodes.

The nitrogen (0001) planes in GaN are arranged similar to the oxygen planes in iron oxides (Fig. 2). As will be shown later these planes get aligned (O-O and N-N triangle sides parallel to each other) in the $Fe_xO_y$ / GaN system resulting in lateral substrate-film mismatch of approximately 7.5 %. For better understanding the orientation of the oxygen and nitrogen triangles (within a single anion plane) relative to the crystallographic axes is shown in Fig. 2 (c). The averaged orientations and lengths of the in-plane O-O and N-N vectors are as follows: |1 0 0|=3.19 Å in GaN, |1/3 -1/3 0|=2.91 Å in α-$Fe_2O_3$, |1/4 -1/4 0|=2.97 Å in $Fe_3O_4$, |1/4 -1/4 0|=2.95 Å in γ-$Fe_2O_3$ and |0 1/3 0|=2.93 Å in ε-$Fe_2O_3$.

## C. Lattice structure of iron oxides and GaN - reciprocal space

Assuming the oxygen and nitrogen planes are aligned as described in the previous section, the iron oxide phases can be readily identified at the growth stage by carrying out RHEED reciprocal space mapping and analyzing the three orthogonal reciprocal lattice views as shown in Fig. 3.

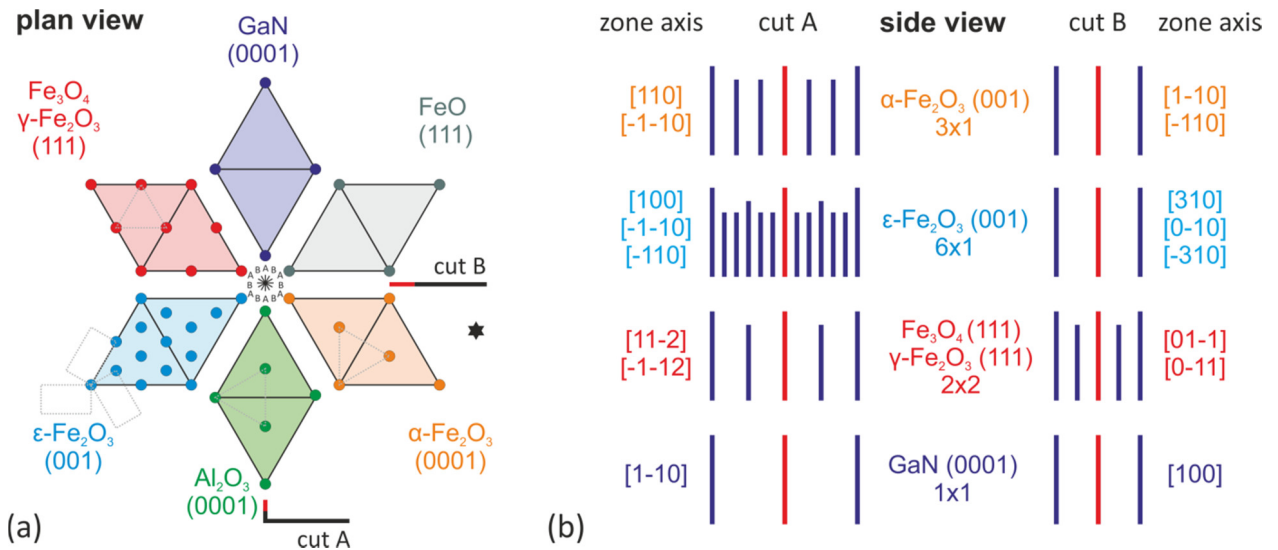

FIG. 3. In-plane epitaxial relations and reciprocal lattice matching between $Al_2O_3$, GaN and iron oxides. The reciprocal space plan view projection (a) shows a pronounced similarity between the in-plane periodicities of the heterostructure components explaining the expected streak patterns to be observed in the two high symmetry reciprocal space zones named cuts A and B. The streak patterns provide a reliable way to distinguish between the cubic (2×2), trigonal (3×1) and orthorhombic (6×1) iron oxide phases (b).

The "plan view" represents the reciprocal space projection onto the substrate surface. The two orthogonal "side-view" reciprocal space cuts (zones) are carried out perpendicular to the substrate surface and pass through the origin. Dealing with the four different crystal structures it is convenient to address the cuts as "cut A" - parallel and "cut B" perpendicular to the O-O (N-N) triangle sides (Fig. 2 (c)). Fig. 3 shows the sketch of the three orthogonal reciprocal space views with "cut A" and "cut B" directions marked on the plan view. Taking into account the low out-of-plane resolution of RHEED operating at a grazing incidence to the flat surface, the sketch of the side-view cuts in Fig. 3 (b) shows streaks in place of Bragg reflections. The experimentally observed Bragg reflections as will be shown later do really get elongated perpendicular to the surface.

The 1×1 in-plane periodicity of the GaN(0001) surface is defined by the atom arrangement in a single nitrogen plane. The iron oxide lattices are approximately commensurate to GaN but have few times larger in-plane periodicities (except for FeO) due to oxygen sublattice distortion and particular iron distribution. This gives rise to N×M streak patterns: 3×1 for $α-Fe_2O_3$, 2×2 for $Fe_3O_4$ / $γ-Fe_2O_3$ and 6×1 for $ε-Fe_2O_3$ (Fig. 3 (a)). Taking into account the higher symmetry of the GaN (0001) surface compared to that of the iron oxides, multiple symmetry related domains are expected – three domains at 120 deg to each other for $ε-Fe_2O_3$ and two domains at 180 deg to each other for $α-Fe_2O_3$, $γ-Fe_2O_3$ and $Fe_3O_4$. In the following sections the experimentally obtained RHEED maps will be shown to fully comply with the sketch in Fig. 3.

## D. RHEED 3D mapping of clean GaN surface and iron oxide transition layer

The three orthogonal reciprocal space maps of the initial GaN(0001) surface are shown in Fig. 4 with the superimposed modeled reflection positions. Here and below the modeled reflections are shown only on half of the map to provide a non-obscured view of the experimental data. The GaN reflections with the low out-of-plane momentum transfer are considerably elongated perpendicular to the surface. This is indicative of total external reflection experienced by electrons at a grazing incidence to a flat step-and-terrace substrate surface (see AFM morphology images below). The horizontal lines present in the cuts A and B are traces of Kikuchi lines that appear very bright for the clean GaN surface.

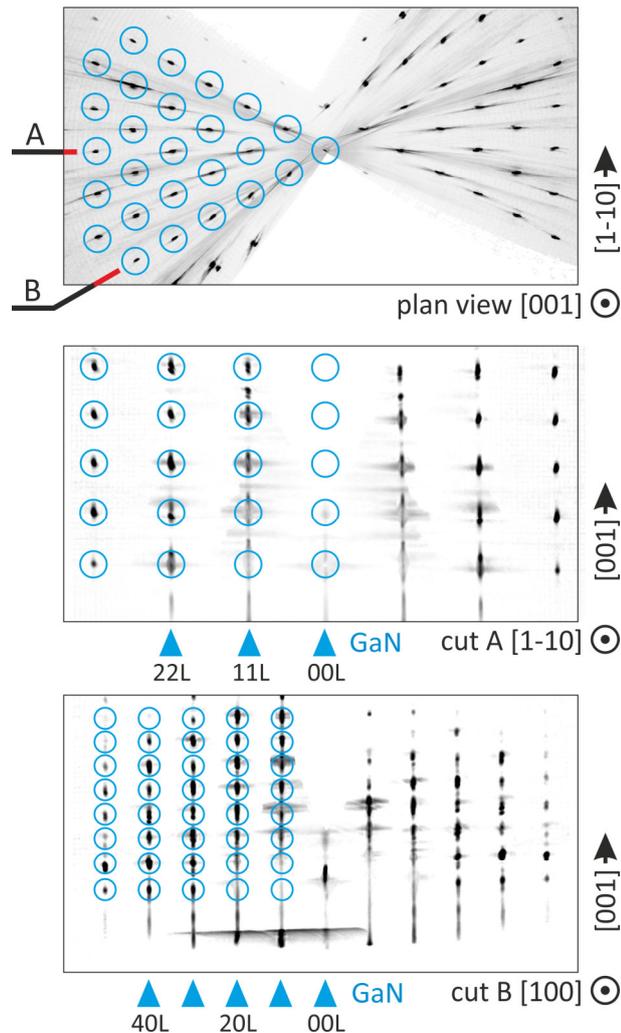

FIG. 4. RHEED reciprocal space map of the clean GaN (0001) substrate prior to the iron oxide deposition. Three orthogonal views including the plan view projection and the two side view reciprocal space cuts A and B are shown in the same scale. The circles represent the modeled GaN reflections. The triangles mark lateral positions of the GaN streaks providing a convenient coordinate system for further examination of the iron oxide reciprocal lattice structure. Distance between GaN [20L] and [00L] is 0.724 Å$^{-1}$.

Upon deposition of about 1 nm of iron oxide the GaN 1×1 map is gradually replaced by a slightly different 1×1 map with extra faint half order streaks visible in reciprocal space cut A (Fig. 5). As indicated by arrows the distance between integer streaks gets slightly larger than in GaN corresponding well to the in-plane periodicity being reduced from 3.19 Å of GaN to ~2.9 Å typical for the iron oxides.

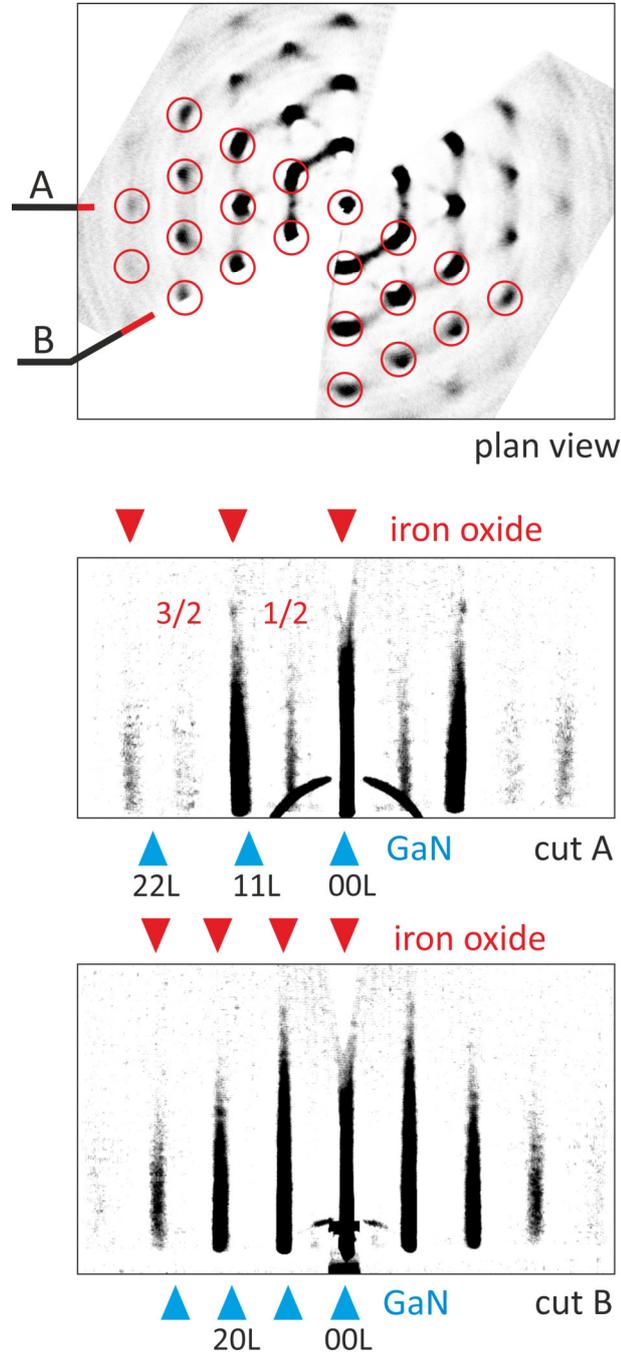

FIG. 5. RHEED reciprocal space map of the iron oxide transition layer. Three orthogonal views including the plan view projection and the two side view reciprocal space cuts A and B are shown in the same scale. Positions of the overgrown GaN streaks are marked to emphasize the slight difference of GaN and iron oxide in-plane periodicities. Distance between GaN [20L] and [00L] streaks shown for scale with blue triangles is 0.724 Å$^{-1}$.

As derived from the previous section, neither 1×1 nor 1×2 in-plane periodicities are characteristic of the $Fe_2O_3$ / $Fe_3O_4$ iron oxides. Presumably at this stage a transition layer is formed with no resemblance to any particular iron oxide phase due to a large number of stacking faults and antiphase boundaries inevitably occurring when a film has a lower surface symmetry and a larger surface cell compared to the substrate. Interestingly the plan view projection shows that the half-order streaks labeled as 1/2 and 3/2 in the cut-A maps are not true 1D streaks but rather the 2D diffuse scattering "walls" connecting the true streaks. Such "walls" are present because the 1×1 streaks experience anisotropic in-plane widening due to the lack of order in the direction perpendicular to the O-O (N-N) in-plane vectors.

According to the RHEED maps the only well pronounced in-plane periodicity present in the transition layer is that of the hexagonal arrangement of atoms within oxygen planes (like in FeO). The out-of-plane periodicity is difficult to extract from the low modulated streaks, but it still can be derived from RHEED specular beam intensity oscillations observed during transition layer nucleation (Fig. 6).

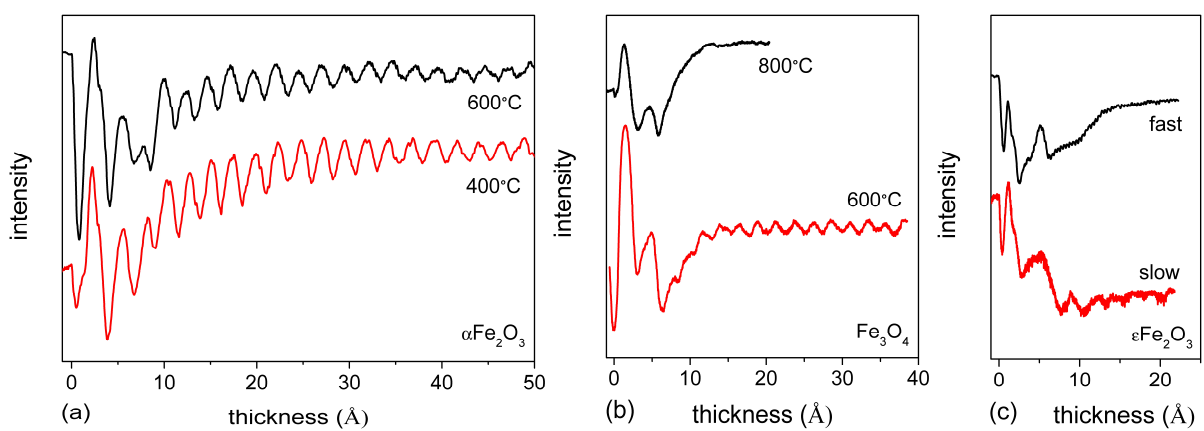

FIG. *6. RHEED specular beam intensity oscillations observed at the beginning of α-$Fe_2O_3$ (a), $Fe_3O_4$ (b) and ε-$Fe_2O_3$ (c) deposition. The most pronounced oscillations corresponding to the layer by layer growth are observed at 600°C and below. The oscillation period after stabilization is approximately 2.4 Å corresponding to a single O-Fe bilayer.*

The first few oscillations show a not easily interpreted shape as it often happens in heteroepitaxy. The reason for this is that during first monolayer nucleation too many parameters are changed at the same time (structure factor of the top layer, refraction of the e-beam, terrace width and surface roughness). It may be however claimed that independent on the growth conditions the ~2.4 Å oscillation period roughly corresponds to the O-Fe single bilayer (like in α-$Fe_2O_3$ and FeO) rather than to the O-Fe-O-Fe double bilayer (like in $Fe_3O_4$ and ε-$Fe_2O_3$). The most pronounced oscillations are observed during nucleation and growth of α-$Fe_2O_3$ in 0.02 mbar of oxygen (Fig. 6 (a)) at 400°C (lowest damping) and 600°C (slightly higher damping). Such behavior corresponds to the layer by layer growth with slow surface roughening. At growth conditions favorable for nucleation of $Fe_3O_4$ (0.02-02 mbar of nitrogen at 600-800°C) and ε-$Fe_2O_3$ (0.2 mbar of oxygen at 800°C) only few pronounced oscillations are observed at the beginning of deposition. The much weaker oscillations are often present afterwards being more pronounced at lower substrate temperature and lower growth rate. The observed behavior corresponds to the Stranski–Krastanov scenario in which the first few monolayers are grown in a layer by layer mode after which the growth switches to island nucleation regime. It is usually at that moment that the intensity stops oscillating and the additional features of particular iron oxide gradually appear. To completely stabilize the particular iron oxide phase with a well recognizable diffraction pattern one needs to grow film having thickness over 5-8 nm. In the next section the 3D reciprocal space maps of even thicker

20-40 nm α-Fe$_2$O$_3$, Fe$_3$O$_4$, γ-Fe$_3$O$_4$ and ε-Fe$_2$O$_3$ films on GaN will be discussed in more details. Comparison to the modeled reciprocal lattices allows phase identification, spotting all the symmetry related domains and confirming the expected epitaxial relations.

**E. RHEED 3D mapping of α-Fe$_2$O$_3$ layer (hematite)**

Depositing Fe$_2$O$_3$ onto GaN(0001) at 400-600°C in 0.02 mbar of oxygen the streak arrangement is gradually transformed from the 1×1 pattern of the transition layer to the √3×√3 R30 pattern of the α-Fe$_2$O$_3$ phase (Fig. 7). Depending on the growth rate and temperature the α-Fe$_2$O$_3$ pattern starts to appear after 5-8 nm of deposition.

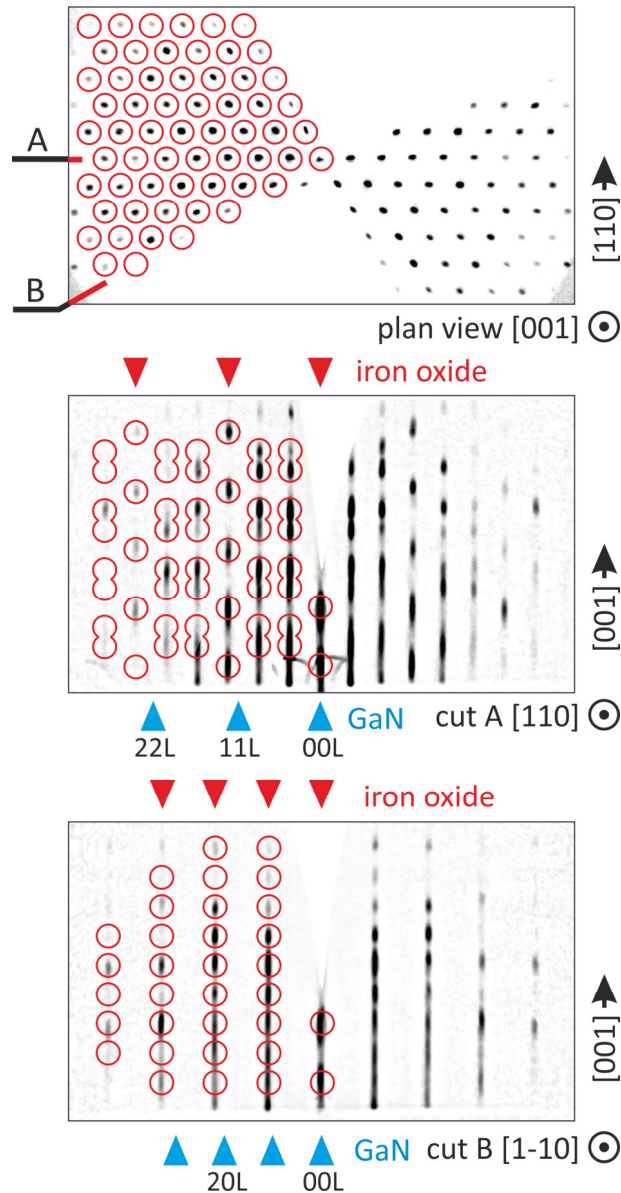

FIG. 7. 3D RHEED reciprocal space maps of α-Fe$_2$O$_3$ layer. Three orthogonal views including the plan view projection and the two side view reciprocal space cuts A and B are shown in the same scale. The circles represent the modeled α-Fe$_2$O$_3$ reflections. Distance between GaN [20L] and [00L] streaks shown for scale with blue triangles is 0.724 Å$^{-1}$.

The observed epitaxial relations are in agreement with those described above (assuming mutual alignment of the O-O and N-N triangles): out-of-plane: GaN[001] || α-Fe$_2$O$_3$ [001]; in-plane: GaN[1-10] || α-Fe$_2$O$_3$ [110] or [-1-10]. The two-fold ambiguity of the in-plane epitaxial relations appears due to the symmetry reasons as the two equally probable orientations may occur when trigonal α-Fe$_2$O$_3$ lattice is placed over the hexagonal GaN lattice. The appearance of the √3×√3 R30 set of streaks is readily visible upon comparing the plan-view in Fig. 7 to the plan view of the initial GaN surface in Fig. 5. Using the side views the αFe$_2$O$_3$ phase can be unmistakably identified by the presence of N/3 streaks in the reciprocal space "cut A" and no additional features in the reciprocal space "cut B". Being first weaker that the integer ones the N/3 streaks fully develop after about 10 nm of deposition at which moment they also get modulated in full correspondence with the modeled α-Fe$_2$O$_3$ reciprocal lattice structure (see circles in Fig. 7). The resulting hematite films are insulating, non-magnetic and have red-orange tint.

**F. RHEED 3D mapping of Fe$_3$O$_4$ and γ-Fe$_3$O$_4$ layers (magnetite and maghemite)**

Depositing Fe$_2$O$_3$ at 600-800°C in 0.02 – 0.2 mbar of nitrogen was found to stabilize the Fe$_3$O$_4$ phase. Already after 2-3 nm of deposition the 1×1 streak pattern of the transition layer starts transforming to the 2×2 streak pattern. The latter is characteristic for the cubic lattice of Fe$_3$O$_4$ (or γ-Fe$_2$O$_3$) with the following epitaxial relations: out-of-plane: GaN[001] || Fe$_3$O$_4$ [111]; in-plane: GaN[1-10] || Fe$_3$O$_4$ [11-2] or [-1-12]. The two-fold ambiguity of the in-plane epitaxial relations emerges due to symmetry reasons as the cubic iron oxide is placed over hexagonal GaN. The half streaks develop in both "cut A" and "cut B" maps gradually getting modulated in full resemblance of the expected Bragg reflection positions of Fe$_3$O$_4$ or γ-Fe$_2$O$_3$ (Fig. 8).

From electron diffraction maps it is almost impossible to distinguish Fe$_3$O$_4$ from γ-Fe$_2$O$_3$ as both oxides obtain the same lattice structure and almost the same lattice constant. The films grown in nitrogen are identified as magnetite (or at least as magnetite-rich) because they have distinct gray metallic tint and are conducting at room temperature. As will be shown below both these properties are not observed in the γ-Fe$_2$O$_3$ films. It is important to note that substitution of Fe$^{III}$ ions with Fe$^{II}$ must be the direct consequence of an increased Fe:O ratio in the plume. The latter can hardly be due to the increased amount of iron but more likely due to the reduced amount of oxygen. Interestingly growing in oxygen deficient (reducing) UHV conditions does not lead to Fe$_3$O$_4$ stabilization. One needs to use nitrogen background atmosphere (argon would not work) at a pressure no less than 0.02 mbar to stabilize magnetite, otherwise α-Fe$_2$O$_3$ would eventually grow. Increasing nitrogen pressure to 0.2 mbar leads to faster magnetite nucleation and subsequently brighter diffraction patterns. The probable explanation could be that nitrogen acts as a sort of oxygen absorber forming NOx compounds. Interestingly the same approach of using nitrogen background atmosphere (to be soon reported elsewhere by the authors) has proved efficient to make Fe$_3$O$_4$ grow on MgO(001) whereas in oxygen one usually gets γ-Fe$_2$O$_3$ as the main growing phase. The observed Fe$_3$O$_4$ / GaN epitaxial relations are in agreement with those discussed in Ref. [29] where Fe$_3$O$_4$ films on GaN were obtained by oxidation of Fe / GaN layers. Apparently iron oxidation technology leads to formation of a partially disordered magnetite film as in the diffraction patterns presented in Ref. [29] the half order streaks are considerably weaker than the integer ones. In contrast to this our magnetite films are fully ordered as confirmed by uniform streak brightness.

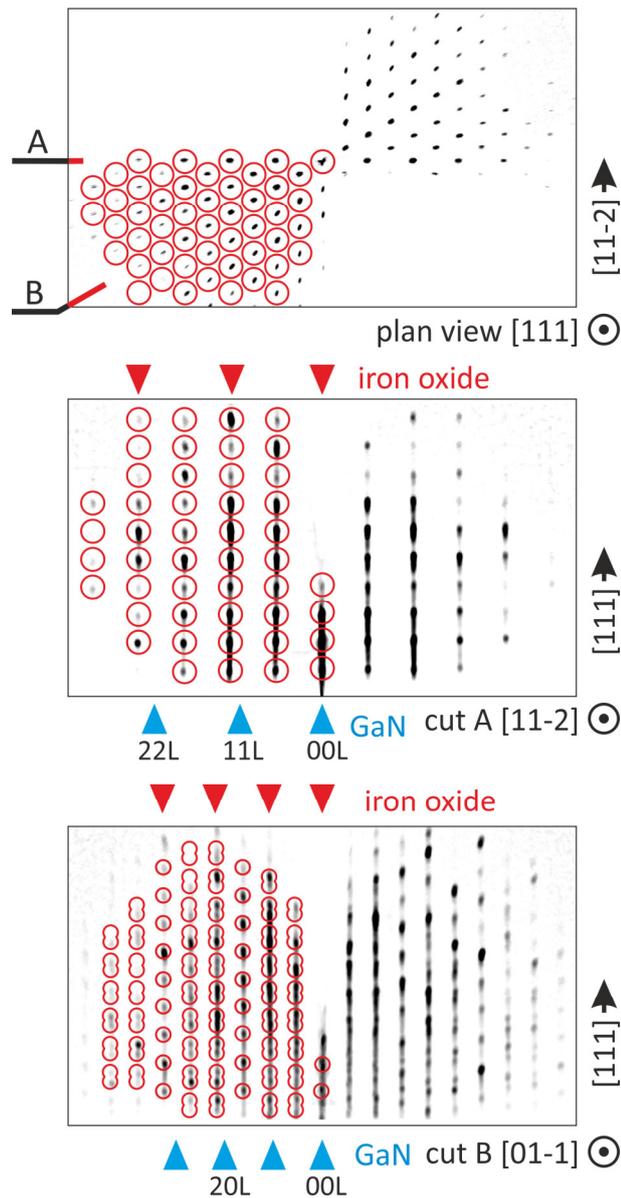

FIG. 8. 3D RHEED reciprocal space maps of $Fe_3O_4$ layer. Three orthogonal views including the plan view projection and the two side view reciprocal space cuts A and B are shown in the same scale. The circles represent the modeled $Fe_3O_4$ reflections. Distance between GaN [20L] and [00L] streaks shown for scale with blue triangles is 0.724 Å$^{-1}$.

The $Fe_3O_4$ – $\gamma$-$Fe_2O_3$ transformation in films grown on GaN affects the full volume when the film thickness is below 20 nm. Therefore to grow thicker maghemite layers, the growth / oxidation procedure has to be repeated multiple times. It is noteworthy that the magnetite – maghemite transformation is expected to proceed through creation of iron octahedral vacancies, that is, iron must travel out of the film volume to meet oxygen at the surface. This process would involve not only volume transformation but also crystallization of extra material at the surface. Though the lattice structure and epitaxial relations do not change upon oxidation a certain amount of surface roughening upon $Fe_3O_4$ oxidation is observed by AFM.

### G. RHEED 3D mapping of ε-$Fe_2O_3$ layers

The most interesting result of the present investigation is the epitaxial stabilization of metastable ε$Fe_2O_3$ phase on a GaN surface. The ε-$Fe_2O_3$ polymorph was shown to grow on GaN in 0.2 mbar of oxygen at 800°C. The diffraction pattern characteristic for ε$Fe_2O_3$ appears after about 5 nm of deposition and stays stable up to the maximum tested thickness of 40 nm. The unmistakable characteristic of εFe$_2$O$_3$ phase is the appearance of reflections on the N/6 streaks in the cut-A reciprocal space maps (Fig. 9). The following epitaxial relations are observed. Out-of-plane: GaN[001] || ε-Fe$_2$O$_3$ [001]; in-plane: GaN[1-10] || ε-Fe$_2$O$_3$ [100] or [-1-10] or [-110]. The three-fold ambiguity of the in-plane epitaxial relations appears due to symmetry reasons when placing orthorhombic ε-Fe$_2$O$_3$ lattice over hexagonal GaN lattice. Noteworthy the "b" lattice parameter of ε-Fe$_2$O$_3$ is √3 times larger than the "a" lattice parameter. This accounts for the coincidence of a great deal of reflections upon 120 deg rotation around the surface normal. The ε-Fe$_2$O$_3$ films are ochre and insulating.

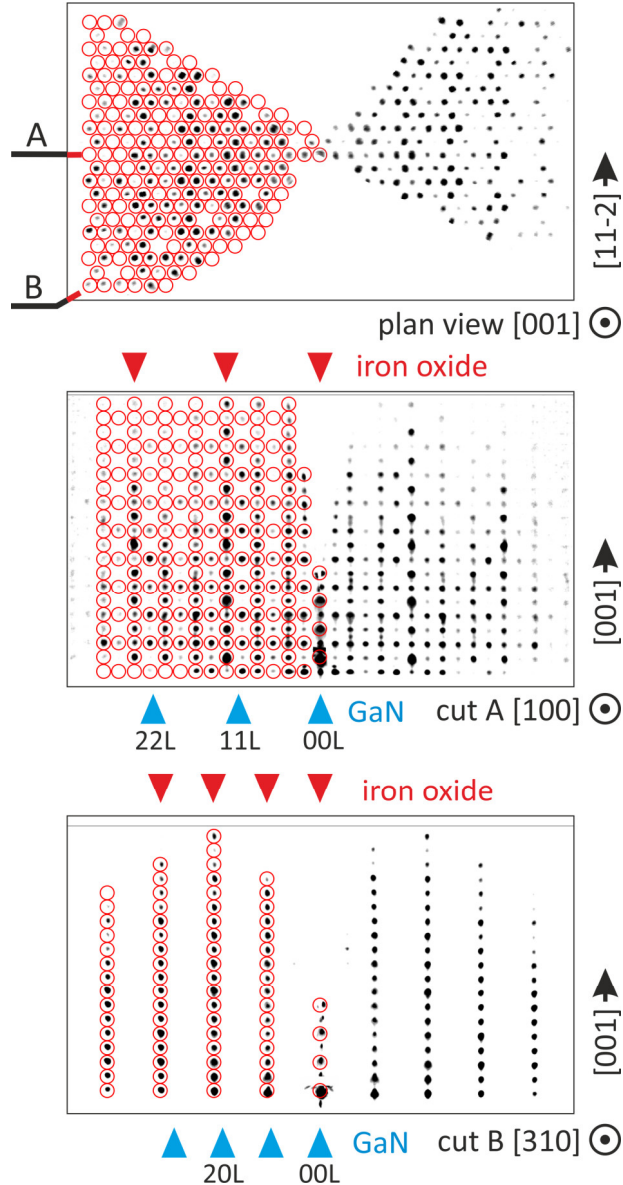

FIG. 9. *3D RHEED reciprocal space map measured for ε-Fe$_2$O$_3$ layer. Three orthogonal views including the plan view projection and the two side view reciprocal space cuts A and B are shown in the same scale. The circles represent the modeled ε-Fe$_2$O$_3$ reflections. Distance between GaN [20L] and [00L] streaks shown for scale with blue triangles is 0.724 Å$^{-1}$.*

## IV. X-RAY DIFFRACTION STUDIES

While RHEED has an excellent surface sensitivity it is in general less precise than XRD in solving lattice structures due to dynamic effects and often lower mechanical accuracy of the in-vacuum goniometers. Moreover since RHEED probes only a thin near surface region it has a limited out-of-plane resolution and cannot rule out recrystallization effects possibly occurring deep inside the film during or after growth. To accurately evaluate the out-of-plane interlayer spacings present in the studied system we have carried out post growth X-ray diffraction analysis of the intensity distribution along the specular crystal truncation rods. Fig. 10 shows such intensity profiles measured in 40 nm films of the four different iron oxides grown on GaN. The intensity is given as a function of momentum transfer perpendicular to the surface ($Q_z$) that has been corrected for possible misalignment using the $Al_2O_3$ and GaN reflections as a reference. The observed iron oxide reflections are in good agreement with the epitaxial relations observed earlier by RHEED.

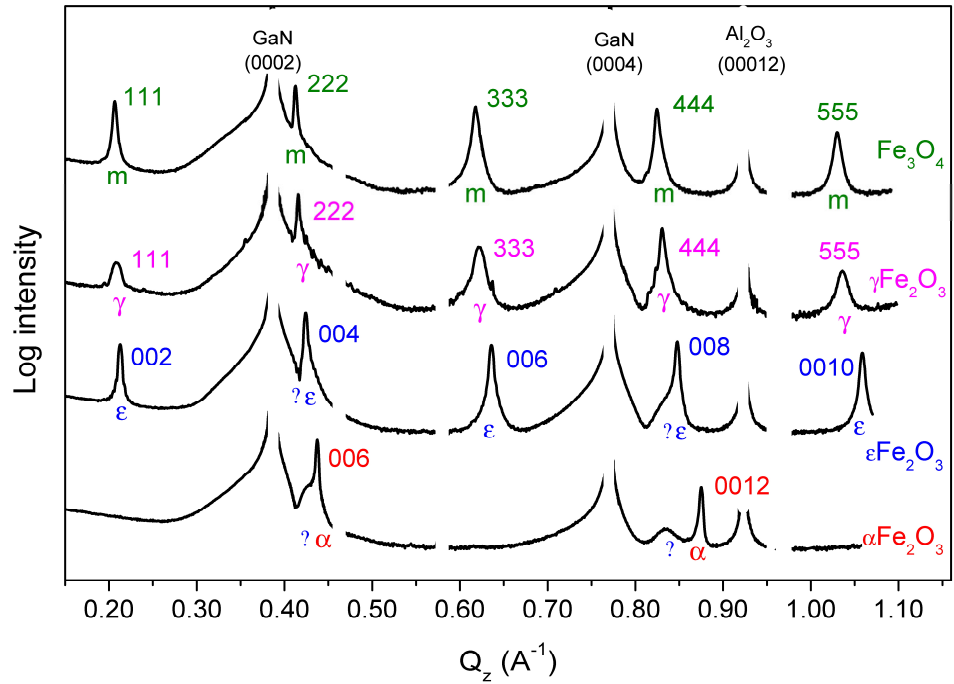

FIG. 10. XRD specular intensity profiles measured in $Fe_3O_4$, $\gamma$-$Fe_2O_3$, $\varepsilon$-$Fe_2O_3$ and $\alpha$-$Fe_2O_3$ epitaxial layers on GaN(0001). Apart from the brightest peaks belonging to GaN and $Al_2O_3$ distinct Bragg reflections of the corresponding iron oxide phases are present. In $\varepsilon$-$Fe_2O_3$ and $\alpha$-$Fe_2O_3$ samples additional two peaks corresponding to the transition layer are clearly visible (labeled with question mark).

An accurate evaluation of the out-of-plane lattice periodicity has been done from the XRD profiles. As was discussed earlier this periodicity is mainly defined by the distance between the oxygen planes in the iron oxide lattice structure and is phase dependent due to different arrangement of the $O_h$ and $T_d$ iron sites as well as the different stacking order of the oxygen planes. Moreover in $\gamma$-$Fe_2O_3$, $Fe_3O_4$ and $\varepsilon$-$Fe_2O_3$ the monolayer contains two oxygen planes leading to twice higher reflection frequency than in $\alpha$-$Fe_2O_3$ in which the monolayer contains only one oxygen plane. It follows from the shown profiles that the $Fe_3O_4$ and $\gamma$-$Fe_2O_3$ (111) layers are slightly expanded perpendicular to the surface while the $\alpha$-$Fe_2O_3$ (0001) and $\varepsilon$-$Fe_2O_3$ (001) layers are slightly compressed. Below are listed the measured distances between the oxygen planes and the corresponding lattice parameters together with given in parentheses bulk values. The lattice parameters for $Fe_3O_4$ and $\gamma$-$Fe_2O_3$ are calculated assuming that their lattice remains cubic.

$Fe_3O_4$:     d=2.428 Å (2.424 Å + 0.1 %)    a=2√3d=8.41 Å (8.398 Å)
$\gamma$-$Fe_2O_3$:     d=2.411 Å (2.405 Å + 0.3 %)   a=2√3d=8.354 Å (8.33 Å)
trans. layer:     d=2.395 Å
$\varepsilon$-$Fe_2O_3$:     d=2.360 Å (2.368 Å - 0.3 %)    c=4d=9.438 Å (9.471 Å)
$\alpha$-$Fe_2O_3$:     d=2.285 Å (2.291 Å - 0.3 %)    c=6d=13.708 Å (13.747 Å)

Interestingly the two extra non-hematite peaks (labeled with question marks in Fig. 10) are present on the α-$Fe_2O_3$ specular profile. These peaks are much wider and weaker than the main α-$Fe_2O_3$ reflections. Their contribution was shown to decrease with film thickness indicating that they originate from a thin layer located at the GaN interface. Most likely this is the fingerprint of the transition layer discussed earlier in this paper. The interlayer distance corresponds to 2.395 Å which is approximately the distance between oxygen planes in undistorted γ-$Fe_2O_3$. The transition layer peaks are very well distinguished in the α-$Fe_2O_3$ layer. The peak width corresponds to a 5 nm film in good agreement with RHEED observations showing the onset of the α-$Fe_2O_3$ phase nucleation after 5.5 nm of deposition. In ε-$Fe_2O_3$ layers a small hump on the low-q side of the (008) reflection is approximately at the same place and of the same width as the transition layer reflection in α-$Fe_2O_3$. In $Fe_3O_4$ and γ-$Fe_2O_3$ the transition layer peak cannot be distinguished likely because this layer is few times thinner than in α-$Fe_2O_3$. In agreement with the period of RHEED oscillations the periodicity of the transition layer corresponds to single rather than double O monolayer. Interestingly the transition layer Bragg peaks show destructive interference with the high-$Q_z$ slopes of the GaN (0002) and GaN (0004) reflections making them highly asymmetrical. The observed coherence between crystal truncation rods of the substrate and the layer indicate that there is a sharp interface between them. This kind of interference often results in noticeable shift of the film peak maximum and can be used to estimate the layer spacing at the interface. Further study of the nature of the transition layer assumed responsible for the nucleation of the exotic ε-$Fe_2O_3$ phase goes beyond the scope of this work and will be highlighted in a separate publication. To conclude, it becomes clear from the XRD studies that the iron oxide films are single phase with out-of-plane interlayer spacing slightly modified with respect to the corresponding bulk values. No post growth phase conversion has been observed so far. Finally the approximately 5 nm thick transition layer present in α- and ε-$Fe_2O_3$ films has sharp interfaces and the interlayer spacing noticeably different from the known iron oxides.

**V. X-RAY ABSORPTION SPECTROSCOPY AND MAGNETIC CIRCULAR DICHROISM**

X-ray absorption and X-ray magnetic circular dichroism have been applied in this work to probe oxidation states and coordination of the Fe atoms in ferrimagnetically ordered sublattices of the studied iron oxides. While the diffraction techniques are sensitive to the crystal structure, the soft X-ray spectroscopy provides the way to prove that the iron atoms are in the expected chemical environment. Most important is that the method is sensible to the non-crystalline fractions (e.g. segregated metallic iron) that cannot be observed with diffraction techniques. The analysis of electronic and magnetic structure is done in this section through comparison to the known absorption spectra of the corresponding bulk materials.

The $L_{23}$ spectra of transition metals correspond to the dipole-allowed 2p - 3d transitions and consist of $L_3$ and $L_2$ main peaks resulting from the spin-orbit coupling. In iron oxides the ligand field of the oxygen atoms splits the Fe 3d into $e_g$ and $t_{2g}$ orbitals providing a highly sensitive tool to probe the coordination environment and the magnetization of individual sublattices. The Fe $L_{23}$ XAS spectra measured in the iron oxide films grown on GaN are shown in Fig. 11 together with the reference spectra adopted from [11,36,37]. In all the studied samples the $L_3$ peak exhibits two major components: a larger peak at 709.5 eV and a smaller satellite at 708 eV. This spectral shape is characteristic of the oxidized iron as opposed to the metallic one [38].

In thick 40 nm α-$Fe_2O_3$ film the splitting is most pronounced originating from the pure octahedral coordination of $Fe^{III}$ ions in hematite [36,38,39]. In 40 nm γ-$Fe_2O_3$ and ε-$Fe_2O_3$ films the $L_3$ splitting is less pronounced in agreement with the earlier studies of these materials: γ-$Fe_2O_3$ [36,39,40] and ε-$Fe_2O_3$ [11]. While the separation between the 708 eV and 709.5 eV components is the same as in α-$Fe_2O_3$, the intensity drop between the peaks becomes less pronounced due to the presence of a Fe$^{III}$ T$_d$ peak right in the middle between the two O$_h$ peaks. Interestingly the same shape of L$_3$ edge as in γ-Fe$_2$O$_3$ and ε-Fe$_2$O$_3$ is observed in the 5 nm transition layer (preceding α-Fe$_2$O$_3$ formation) indicating the presence of T$_d$ sites near the interface. The non-hematite nature of the transition layer is in agreement with our diffraction data presented earlier in this paper.

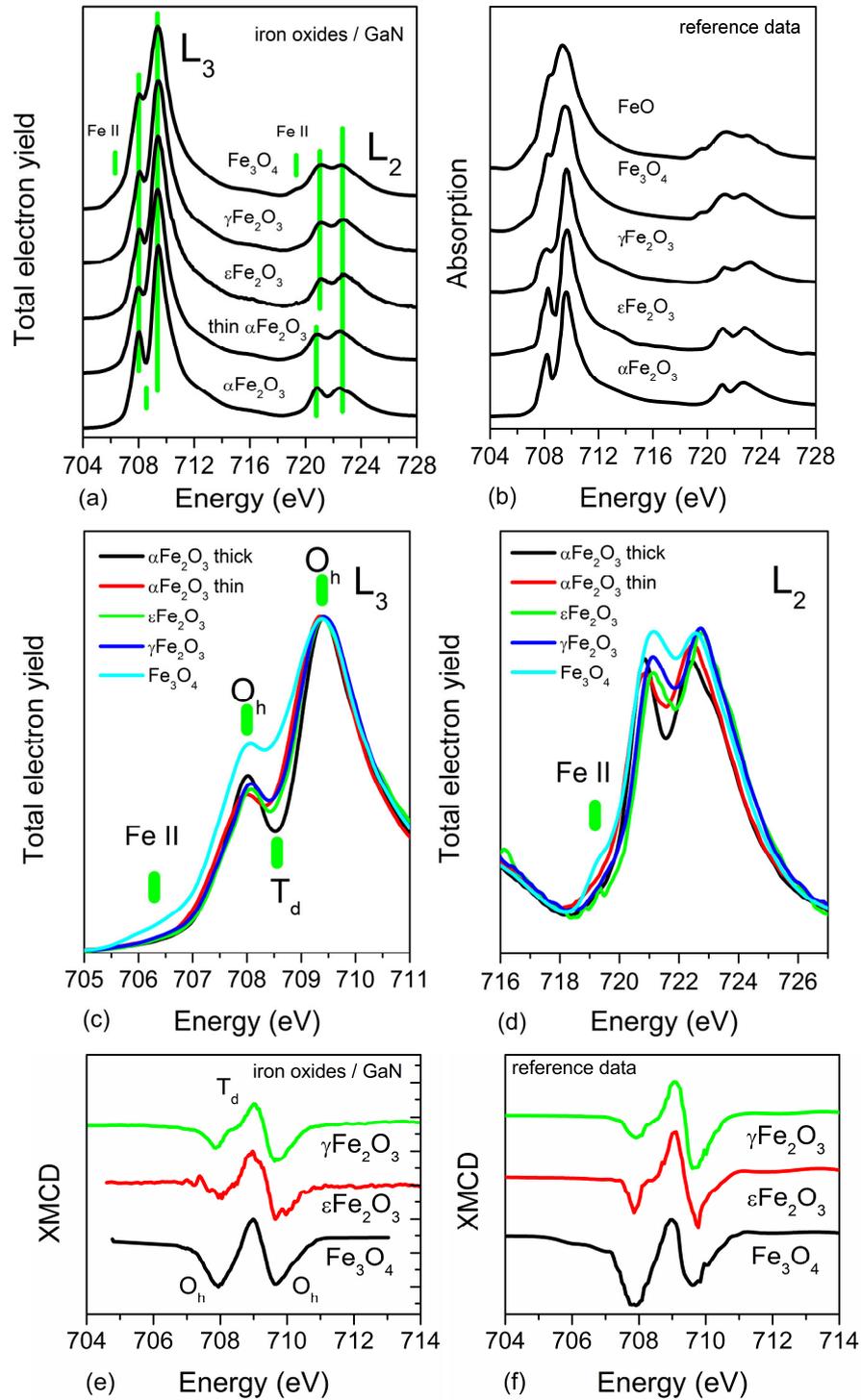

FIG. 11. X-ray absorption spectra obtained from the iron oxide epitaxial films on GaN showing the full L$_{23}$ spectral range (a) and the detailed L$_3$ and L$_2$ spectra (c, d). Reference data adopted from [11,36,37] is shown for comparison (b). The features in the spectra belonging to Fe II and Fe III as well as to the octahedral (Oh) and tetrahedral (Td) iron sites are marked. XMCD spectra obtained from the iron oxide epitaxial films on GaN (e). Reference data adopted from [11,36,37] is shown for comparison (f).

It is noteworthy that in the only existing paper [11] describing X-ray absorption in ε-Fe$_2$O$_3$, the spectrum is more like in α-Fe$_2$O$_3$ with high contrast L$_3$ splitting (Fig. 11 (b)) while our data show resemblance to γ-Fe$_2$O$_3$. The latter seems more natural as unlike in α-Fe$_2$O$_3$ in ε-Fe$_2$O$_3$ the T$_d$ iron sublattice is known to exist in addition to the three distorted O$_h$ sublattices. In contrast to the pure trivalent oxides, our Fe$_3$O$_4$ films exhibit an extra shoulder at 706.5 eV and a considerably higher satellite at 709.5 eV. These features are known to be characteristic of O$_h$ Fe$^{II}$ sites in magnetite [38–43]. The shape of the L$_2$ peaks in the studied iron oxide films bears some similarity to the L$_3$ peak exhibiting the Fe$^{II}$ low energy shoulder in Fe$_3$O$_4$ and a better resolved satellite in α-Fe$_2$O$_3$.

The XMCD measurements at Fe L$_{23}$ edge have been carried out to investigate magnetic nature of Fe sublattices making comparison to the corresponding bulk materials. Within the experiment accuracy no dichroic signal has been detected in the expected to be weak ferrimagnetic α-Fe$_2$O$_3$ films. The XMCD spectra of the other three iron oxides compared to the reference spectra are shown in Figs. 11 (e, f). The general trend is that the L$_3$ XMCD spectra are dominated by two negative peaks corresponding to the Fe magnetic moments in O$_h$ sites aligned parallel to the field and a positive peak in between the O$_h$ peaks corresponding to the Fe magnetic moments in T$_d$ site aligned antiparallel to the field. In the pure Fe$^{III}$ oxides (γ-Fe$_2$O$_3$, ε-Fe$_2$O$_3$) the higher energy O$_h$ peak is dominant (see reference spectra in Fig. 11 (f) adopted from [11,36]). In the mixed II / III Fe$_3$O$_4$ the low energy peak is higher and wider due to the presence of the O$_h$ Fe$^{II}$ ions [29,44]. The net magnetization in ferrimagnetic iron oxides results from the antiparallel alignment of magnetic moments at the octahedral and tetrahedral Fe sites. In contrast to the XMCD spectra of metallic iron for which the L$_3$ peaks measured at opposite light helicities show different heights, in iron oxides these peaks are shifted in energy.

To summarize, the soft X-ray absorption spectroscopy has confirmed that in the grown films iron atoms reside in the expected oxidation state and crystallographic environment and that the films may be considered chemically pure. For all the cases the distinction can be made between the tetrahedrally and octahedrally coordinated iron atoms both in XAS and XMCD spectra. This makes possible a more detailed XMCD study of individual magnetic sublattices in the ferrimagnetic iron oxides. Interestingly the transition layer was found to be different from αFe$_2$O$_3$ in that it contains tetrahedrally coordinated iron.

## VI. IN-PLANE MAGNETIZATION REVERSAL IN THE IRON OXIDE LAYERS ON GAN

In plane magnetization reversal curves of the iron oxide layers have been investigated by MOKE and VSM to compare the magnetic properties of the grown films to those of the same materials in bulk or nanoscale form. Fig. 12 (a) emphasizes the big difference of the typical M(H) curves measured in Fe$_3$O$_4$, α-Fe$_2$O$_3$, γ-Fe$_2$O$_3$ and ε-Fe$_2$O$_3$ layers.

### A. Epsilon ferrite

In ε-Fe$_2$O$_3$ (001) layers the easy magnetization [100] axis should lie in the film plane accounting for the hard magnetic nature of the in-plane magnetization curves. Such curves have been measured by VSM showing saturation of 110 emu/cc, coercivity of 8 kOe and saturation field of 20 kOe (Fig. 12 (c)). These values are typical for εFe$_2$O$_3$ nanoparticles and films reported by different authors [13,17–22]. Similar to the other reports we observe loops of wasp-waist shape exhibiting abrupt magnetization jumps at zero magnetic field. The wasp waist loop is explained by the presence of magnetically hard and soft components in ε-Fe$_2$O$_3$ film. An example of loop decomposition is shown in Fig. 12 (c). The hard magnetic component loop shape is in agreement with the idea of ε-Fe$_2$O$_3$ film consisting of domains with three possible orientations of the easy magnetization ε-Fe$_2$O$_3$ [100] axis at 120 deg to each other. In this configuration the magnetic field is always at an angle with the easy axis of at least two domains which makes the magnetization loops non-rectangular. The origin of the soft component loop is arguable – different works relate it to the presence of magnetite or maghemite impurities, to the antiphase boundaries or to the magnetically soft behavior of the undistorted $O_h$ sublattices. According to our diffraction measurements the only parasitic phase that can be present to some extent at the near surface region of the not optimally grown ε-Fe$_2$O$_3$ films is hematite. However due to the negligible magnetic moment, hematite cannot account for magnetically soft component that in some ε-Fe$_2$O$_3$ gets as strong as 40 emu/cc in saturation. It is likely that the soft loop is related to the transition layer that is rich in small superparamagnetic ε-Fe$_2$O$_3$ grains forming at the nucleation stage. According to our experiments increasing the transition layer thickness (e.g. by decreasing the growth temperature by 50°C – 100°C) makes the soft loop more pronounced.

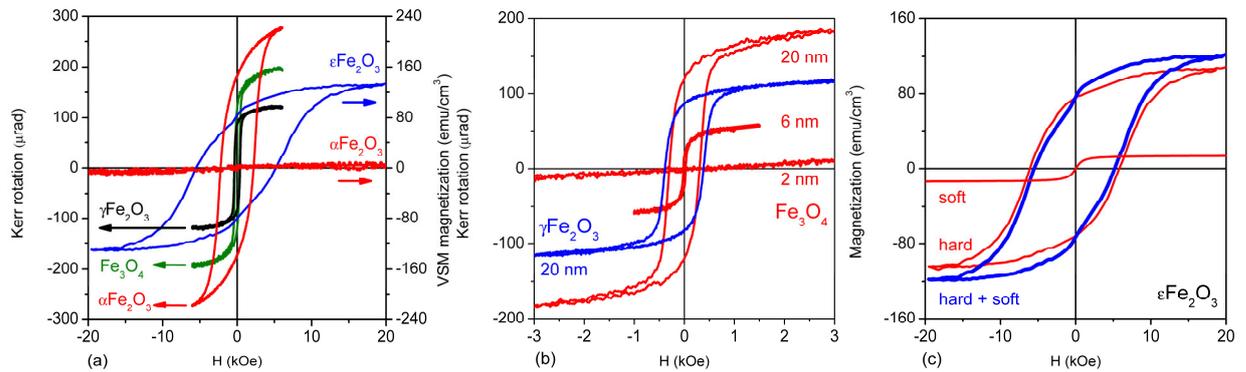

FIG. 12. (A) A comparison chart of typical in-plane magnetization reversal curves measured by longitudinal Kerr effect at λ=405 nm and VSM for α-Fe$_2$O$_3$, Fe$_3$O$_4$, γ-Fe$_2$O$_3$ and ε-Fe$_2$O$_3$ films on GaN. The drastic variation of iron oxide magnetic properties is visible. (B) In-plane MOKE magnetization curves in Fe$_3$O$_4$ films of different thickness and in γ-Fe$_2$O$_3$ showing gradual transition from small might be superparamagnetic to large ferromagnetic grains. (C) In-plane VSM magnetization curve in ε-Fe$_2$O$_3$ film decomposed into magnetically hard and magnetically soft components.

**B. Hematite**

We have also looked for the presence of the weak ferromagnetism in α-Fe$_2$O$_3$ films grown on GaN. At above the Morin transition temperature of 250 K the canted antiferromagnetically coupled magnetic moments in α-Fe$_2$O$_3$ are known to result in spontaneous magnetization lying in the (0001) basal plane [45,46]. As the canting angle is just a fraction of a degree, the resulting magnetic moment is very small - of the order of few emu/cc. Such small magnetization could not be detected in our XMCD experiments. Neither was VSM capable to accurately measure a magnetization loop in a 40 nm α-Fe$_2$O$_3$ film. A very noisy VSM loop in Fig. 15a is shown to emphasize the tiny <5 emu/cc saturation magnetization in α-Fe$_2$O$_3$ compared to the much larger 110 emu/cc value measured for ε-Fe$_2$O$_3$. Interestingly, despite the small magnetic moment a very pronounced Kerr polarization rotation was observed in the studied α-Fe$_2$O$_3$ films (Fig. 12 (a)). The anomalously strong magnetooptical effects in weak ferromagnets are known to occur due to dependence on the antiferromagnetic vector rather than just on the total magnetization [47]. In the very early works linear magnetic birefringence [48] and Kerr effect [49] in hematite have been reported. The Kerr rotation in hematite is claimed to be as strong as in the rare earth iron garnets [50] and can be used to visualize magnetic domains with polarization microscopy [51]. Surprisingly there are no recent works on magnetooptics in hematite films though it appears that the ratio of MOKE to VSM signal at saturation provides a very sensitive tool to distinguish αFe$_2$O$_3$ from other iron oxide phases. The observed MOKE loop shape in our α-Fe$_2$O$_3$ / GaN films is in general agreement with the VSM magnetization curves reported by different authors for bulk samples containing hematite single domain nanoparticles exhibiting coercivity of few kOe and saturation field above 20 kOe [52–54]. Such magnetization behavior is in contrast with the properties of the large natural hematite crystals that have very much lower coercive forces of 3-30 Oe [55]. The main source of the high coercivity in nanoparticles is believed to be the magnetoelastic anisotropy that can be associated with subparticle structure, strain or twinning [56,57].

**C. Magnetite and maghemite**

The MOKE magnetization loops of 20 nm Fe$_3$O$_4$ layers grown in a range of growth conditions (600-800°C in 0.02 to 0.5 mbar of nitrogen) exhibit coercivity of 300-400 Oe and remanence of about 70% of the saturation value (Fig 12 (b)). Very similar magnetization loops have been observed in Fe$_3$O$_4$ films grown by PLD on Si(001) [58]. A similar loop shape with somewhat lower coercivity of 120-150 Oe was observed in Fe$_3$O$_4$ / GaN(0001) films grown by oxidizing a few nm thick iron layer [29]. The Kerr rotation of the γFe$_2$O$_3$ layers is typically 1.5 times weaker compared to that of Fe$_3$O$_4$. With decreasing magnetite layer thickness from 20 nm to 6 nm and further to 2 nm the saturation field gets lower while the remanent magnetization disappears resembling approach to superparamagnetic behavior.

**D. Single vs multiple domains**

The aforementioned thickness dependence for the magnetite films indicates that at nucleation stage the film is not uniform but consists of magnetically non interacting small grains. The high coercive field observed in α-Fe$_2$O$_3$ and ε-Fe$_2$O$_3$ films is also indicative of the single domain particles rather than a multiple domain film. It is well known that the maximum coercivity for a given material occurs within its single domain range where the magnetization reversal is through synchronous magnetic moment rotation of the whole grain. To overcome the magnetocrystalline anisotropy in this case requires more energy than to move the domain wall in a multiple domain film. The critical size for single domain behavior depends on several factors including, the saturation magnetization. It is estimated to be about 80 nm for magnetite [59] and 20-100 micrometers for hematite (20-100 micrometers) [60]. The diameter of the mounds observed in our films by AFM is in the single domain region. One has to take into account also that for very small single domain (superparamagnetic) particles the randomizing effect of thermal energy becomes important making the material magnetically soft. The idea of the film consisting of single crystal columns separated by antiphase boundaries is supported by diffraction data (showing the presence of multiple symmetry related crystallographic orientations) and by the surface morphology studies showing mounds at column tops. The columnar structure is the result of higher symmetry and lower unit cell of the GaN substrate compared to that of the iron oxides. The islands nucleate in phase with GaN but with a phase shift with respect to each other. Upon coalescence an antiphase boundary is formed and the islands keep growing vertically forming columns.

To summarize, the conducted magnetization reversal measurements have proved that the in-plane magnetic behavior in thick iron oxide films grown on GaN is characterized by semi-rectangular magnetization loops with reasonably high values of remanence. The lowest coercivity has been observed in magnetite and maghemite films, while the largest coercivity takes place as expected in the ε-Fe$_2$O$_3$ layers. Unexpectedly high magneto-optic response has been observed in hematite films. A two-component loop in agreement with the other studies is observed in εFe$_2$O$_3$ films. Indications are obtained of that the building blocks of the iron oxide films are single domain magnetic particles. In general the growth of iron oxide films on GaN is proved applicable for creation of magnetic-on-semiconductor heterostructures.

**VII. SUMMARY**

In the present work we have investigated the tunable polymorphism of epitaxial iron oxides on GaN(0001) surface. Four crystallographically and magnetically different iron oxide phases - $Fe_3O_4$, $\gamma$-$Fe_2O_3$, $\varepsilon$-$Fe_2O_3$ and $\alpha$-$Fe_2O_3$ - have been stabilized by means of the single target Laser MBE technology. Fabrication of single phase layers by controllable variation of substrate temperature, buffer gas composition and pressure has been demonstrated. Among the stabilized iron oxide films the most intriguing is the metastable epsilon ferrite $\varepsilon$-$Fe_2O_3$ the growth of which in the form of epitaxial layers has been reported in very few works up to now. The epitaxial control over the iron oxide polymorphism on GaN is claimed feasible because of the particular iron oxide lattice structure consisting of a close packed stack of oxygen planes and iron atoms sandwiched in between these planes. Though the oxygen planes are indexed differently - $Fe_3O_4$ (111), $\gamma$-$Fe_2O_3$ (111), $\varepsilon$-$Fe_2O_3$ (001) and $\alpha$-$Fe_2O_3$ (0001) - their internal slightly distorted hexagonal structure is quite similar. Our RHEED and XRD studies have shown that the iron oxide layers crystallize with the oxygen planes parallel to the GaN (0001) surface and the in-plane epitaxial relations defined by the coincidence of the O-O and N-N vectors within the corresponding planes. With these epitaxial relations the iron oxide phase may be easily switched during the growth in the way similar to the ABC-ABC / AB-AB-AB stacking switching observed in FCC / HCP lattices [61] with the difference that in the case of iron oxides there exist more than two stacking orders. The lower symmetry of the iron oxide planes compared to GaN accounts for multiple possible in-plane orientations – two for $Fe_3O_4$, $\gamma$-$Fe_2O_3$, $\alpha$-$Fe_2O_3$ and three for $\varepsilon$-$Fe_2O_3$. This ambiguity together with the few times larger in-plane lattice periodicity of the iron oxide is supposed to be responsible for the appearance of phase shifted regions during the film nucleation stage. The thick film are shown to be uniform with low surface roughness (as confirmed by AFM) and are supposed consisting of single crystalline columns separated by antiphase boundaries.

An important component of the $Fe_xO_y$ / GaN technology is the transition layer that forms at the initial growth stage. This layer has the out-of-plane periodicity coincident with the single oxygen interplane distance and the in-plane periodicity defined by that of the would-be-ideal hexagonal oxygen plane. The longer in-plane periodicities characteristic of the bulk iron oxides are not present at this stage as shown by diffraction indicating that the transition layer consists of small regions being in phase with GaN but out of phase with each other. The defect rich transition layer is the thickest in $\alpha Fe_2O_3$ and $\varepsilon Fe_2O_3$ films and much thinner in $Fe_3O_4$, and $\gamma$-$Fe_2O_3$ layers. Interestingly the transition layer was shown to contain tetrahedrally coordinated iron.

The observed surface morphology might be the result of the island nucleation and coalescence mechanism. This kind of growth was reported earlier for $\varepsilon Fe_2O_3$ layers on STO and YSZ [13,21,22]. The columnar structure explains well the magnetic behavior of the iron oxide films on GaN suggesting that columns are single magnetic domains. In a single domain system the coercivity is known to be the highest as the magnetocrystalline anisotropy has to be overcome upon simultaneous magnetization rotation of the whole domain. This is well demonstrated for the case of $\varepsilon$-$Fe_2O_3$ films in which coercivity of 10 kOe and saturation field of 20 kOe have been observed. The reason for the hard magnetic behavior is the extremely high magnetocrystalline anisotropy in $\varepsilon$-$Fe_2O_3$. Much softer in-plane magnetization behavior has been observed in $Fe_3O_4$ and $\gamma$-$Fe_2O_3$ films with typical coercivity values of 300-400 Oe in agreement with other works. Decreasing film thickness was shown to result in appearance of superparamagnetic loop shape most likely due to the small size of the single domain particles present at the nucleation stage.

Interestingly the rather hard magnetic behavior was also observed in α-$Fe_2O_3$ films in agreement with various studies of hematite nanoparticles. The very low total magnetization in thin α-$Fe_2O_3$ films makes it difficult to apply VSM. However the rather high magneto-optical effects in hematite made it possible to monitor magnetization reversal in α-$Fe_2O_3$ films by means of MOKE.

The X-ray absorption spectra measured at the L edge of iron were shown to be in good agreement with the reference spectra obtained in a number of works describing different iron oxides in bulk and nanocrystalline forms. This is an important proof of the chemical pureness of the grown films. A careful analysis of the spectral shape allows distinguishing between the octahedral and tetrahedral coordination of iron as well as the iron oxidation state being different for different oxides. The latter observation makes possible further studies of the individual magnetic sublattices in the ferromagnetic iron oxides by means of XMCD. This is especially challenging to be carried out for the scarcely studied epsilon ferrite obtaining four ferrimagnetically ordered iron sublattices.

In general the obtained results related to iron oxide stabilization on GaN(0001) are believed to be helpful in clarifying the influence of technological parameters on the laser MBE process and in getting hold of the mechanisms guiding the particular phase choice during film nucleation and growth. The shown flexibility in phase stabilization by appropriately tuning the Laser MBE technological parameters makes it possible to predict that a similar approach may be used to grow iron oxides on the AlGaN (0001) surface. More complex compounds like $Fe_xGa_yAl_{1-x-y}O_3$ can be also tried for the role of the ferromagnetic layer. Combining epitaxial layers of magnetically ordered materials with a highly suitable for device applications GaN semiconductor surface is supposed to present potential interest for designing novel (opto-) electronic and spintronic devices for room temperature operation.

## *ACKNOWLEDGMENTS*

The authors wish to acknowledge W. V. Lundin for providing GaN / $Al_2O_3$ wafers, the beamline staff at PF for kind assistance in conducting experiments. Beamtime at PF was granted for projects 2014G726, 2014G725 and 2016G684. Support for conducting synchrotron measurements has been received from Nagoya University. This work has been supported by RSF (project no. 17-12-01508) in part related to development of the ferrite-on-GaN growth technology and by Government of the Russian Federation (Program P220, project No.14.B25.31.0025) in part related to sample characterization.